\documentclass[aps,twocolumn,prl,longbibliography,superscriptaddress]{revtex4-2}
\usepackage[colorlinks,linkcolor=black,citecolor=blue,urlcolor=black]{hyperref}
\usepackage{amsmath,amsfonts,amssymb,graphicx,bm,color}

\begin{document}

\author{G. Spada}
\affiliation{
    Pitaevskii BEC Center, CNR-INO and Dipartimento di Fisica, Universit\`a di Trento, I-38123 Trento, Italy}
\affiliation{
	School of Science and Technology, Physics Division, Universit\`a di Camerino, 62032 Camerino, Italy}
\affiliation{
	INFN, Sezione di Perugia, I-06123 Perugia, Italy
}
\author{S. Pilati}
\affiliation{
	School of Science and Technology, Physics Division, Universit\`a di Camerino, 62032 Camerino, Italy}
\affiliation{
	INFN, Sezione di Perugia, I-06123 Perugia, Italy
}
\author{S. Giorgini}
\affiliation{
    Pitaevskii BEC Center, CNR-INO and Dipartimento di Fisica, Universit\`a di Trento, I-38123 Trento, Italy}
\title{Quantum Droplets in Two-Dimensional Bose Mixtures at Finite Temperature}

\begin{abstract}
	We investigate the formation of quantum droplets at finite temperature in attractive Bose mixtures subject to a strong transverse harmonic confinement. By means of exact path-integral Monte Carlo methods we determine the equilibrium density of the gas and the liquid as well as the pressure vs.~volume dependence along isothermal curves. Results for the equation of state and for the gas-liquid coexistence region in quasi-2D configurations are compared with calculations in strictly two dimensions, finding excellent agreement. Within the pure 2D model we explore the relevance of the quantum scale anomaly and we determine the critical interaction strength for the occurrence of the first-order gas to liquid transition. Furthermore, we find that the superfluid response develops suddenly, following the density jump from the gas to the liquid state.
\end{abstract}

\maketitle

\begin{figure*}
	\centering
	\includegraphics[width=0.98\textwidth]{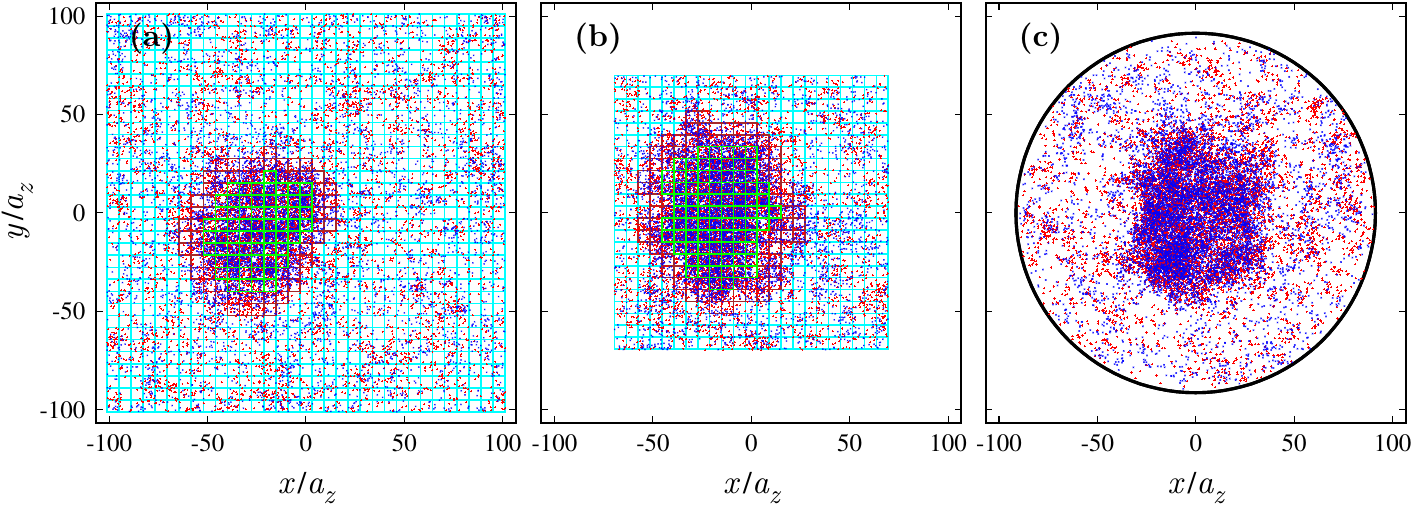}\vspace{-5pt}
	\caption{
		Snapshots of particle positions of component 1 (red) and component 2 (blue) particles in a quasi-2D mixture with harmonic confinement along the $z$ direction projected onto the  $x$-$y$ plane. Panels (a) and (b) refer to a system with periodic boundary conditions along the $x$ and $y$ directions. Panel (c) refers to a mixture subject to a cylindrical hard-wall confinement. The average 2D density is $na_z^2 = 0.4$ [panel (a)], $na_z^2 = 0.8463$ [panel (b)] and $na_z^2 = 0.625$ [panel (c)]. Furthermore, the number of particles is $N_1+N_2=16384$, the temperature is $k_BT/(\hbar\omega_z)=0.2$, and intraspecies, interspecies scattering lengths are $a/a_z=0.03162$ and $a_{12}=-1.1a$, respectively. In panels (a) and (b) the squares indicate tiles identified as liquid region (green), gas region (cyan), and interface region (brown). Notice how the volume reduction from panel (a) to panel (b) results in a larger droplet leaving the average density of both the liquid and the gas unaltered.
	}
	\label{fig1}
\end{figure*}

Attractive mixtures of dilute Bose gases provide a fascinating test bed for novel quantum phenomena, such as droplet formation in a liquidlike state stabilized at ultralow densities, and for the perspectives of new thermodynamic features emerging in the liquid-gas coexistence region and in the critical behavior at the phase transition, including the onset of superfluidity. Quantum droplets in three dimensions (3D) have been observed in homonuclear mixtures of $^{39}$K~\cite{Science.359.301, PhysRevLett.120.235301} and in heteronuclear mixtures of $^{41}$K-$^{87}$Rb~\cite{PhysRevResearch.1.033155}. The evidence of self-bound clusters in these time-of-flight experiments is well explained by a zero-temperature ($T=0$)  energy functional, where repulsive beyond mean-field correlations stabilize the attractive mean-field interactions which would lead to collapse~\cite{PhysRevLett.115.155302}. Numerical simulations based on more microscopic descriptions also confirm this scenario~\cite{PhysRevB.97.140502, PhysRevA.99.023618, PhysRevA.104.033319}. The same stabilizing mechanism appears to be responsible for the formation of other self-bound droplets, realized in single-component dipolar gases~\cite{Schmitt2016, PhysRevX.6.041039}.

Determining the finite-temperature behavior requires accurate computational techniques, since beyond mean-field effects have to be properly accounted for even in the critical region close to the Bose-Einstein condensation (BEC) transition.
For 3D geometries, recent path-integral Monte Carlo (PIMC) simulations~\cite{PhysRevLett.131.173404} predicted that balanced mixtures undergo a first-order transition featuring a discontinuous jump of the density and that BEC occurs discontinuously as the density jumps from the equilibrium value of the gas to the one of the liquid. The first-order transition line is predicted to terminate at a critical point where the BEC transition turns second order with a vanishing condensate fraction.

In fact, such a rich phase diagram was speculated to emerge in attractive single-component Bose gases stabilized by repulsive three-body interactions~\cite{PhysRevA.104.043301} or in quantum fluids interacting with Lennard-Jones potentials when the particle mass is hypothetically tuned below the value of the $^{4}$He mass~\cite{Son_2021,10.1073/pnas.2017646117}. In this regard, attractive Bose mixtures represent a more realistic platform where this novel and intriguing thermodynamic behavior could be observed. However, while the predicted sizable jump of the density at the transition would in principle be detectable, it occurs at densities as large as $n_{3D}\sim10^{15}$ cm$^{-3}$, implying short lifetimes due to three-body collisions thereby making it difficult to observe the liquid droplet in thermal equilibrium with the surrounding gas phase.

In two dimensions (2D) quantum droplets have been predicted to occur at $T=0$, thanks again to quantum fluctuation effects in the ground-state energy, which produce a stabilization of the interspecies attractive interaction~\cite{PhysRevLett.117.100401}. In particular, a relevant role is played by the so-called 2D quantum scale anomaly, emerging as a logarithmic density dependence of the coupling constant which breaks the scaling symmetry and whose signature was observed in 2D superfluids~\cite{doi:10.1126/science.aau4402}.
Furthermore, in 2D a finite-temperature BEC is ruled out by the Mermin-Wagner theorem~\cite{PhysRevLett.17.1307,PhysRev.158.383}, but a superfluid phase still forms below the Berezinskii-Kosterlitz-Thouless (BKT) transition~\cite{Berezinsky:1972rfj,Kosterlitz1973}. This leads one to wonder if and how 2D quantum droplets form at finite temperatures and how a first-order gas-liquid transition would connect with the ordinary BKT superfluid transition.

In this Letter we perform large-scale PIMC simulations to investigate quantum droplet formation in a 2D attractive Bose mixture at finite temperature. First, we consider quasi-2D configurations where microscopic interactions are 3D and the mixture is subject to a strong transverse harmonic confinement. Simulations show the formation of high density clusters in equilibrium with a low density background, with changes in the size of the simulation cell only affecting the relative volume of the two regions, without changing the corresponding densities (see Fig.~\ref{fig1}). In addition, the pressure along isothermal lines shows the typical flattening of a first-order coexistence region followed by a rapid increase as the density rises in the liquid phase. The same features are found with a pure 2D model where interactions are treated via a density dependent coupling constant, which accounts for the quantum scale anomaly. In particular, the presence of this quantum anomaly is crucial to stabilize the liquid state and to reproduce quantitatively the results of quasi-2D simulations. Within the pure 2D model, the weak breaking of scale invariance leads to an only marginal deviation from the universal temperature dependence of the equation of state, and we also analyze the emergence of a critical point as a function of the interspecies coupling where the first order transition terminates. Our results indicate the possibility of observing the rich physics of quantum droplets and liquid-gas transition in 2D attractive Bose mixtures for densities and coupling parameters easily achieved in current experiments. Furthermore, these predictions are universal in terms of the 2D coupling constants.

Our simulations of mixtures with transverse harmonic confinement are based on the following Hamiltonian
\begin{eqnarray}
	H&=&-\frac{\hbar^2}{2m}\sum_{i=1}^{N_1}\left(\nabla_i^2
	-\frac{z_i^2}{a_z^4}\right)
	-\frac{\hbar^2}{2m}\sum_{i^\prime=1}^{N_2}\left(\nabla_{i^\prime}^2-\frac{z_{i^\prime}^2}{a_z^4}\right) \nonumber \\ &+&
	\sum_{i<j} ^{N_1} v(r_{ij}) +
	\sum_{i^\prime<j^\prime}^{N_2} v(r_{i^\prime j^\prime})  + \sum_{i,i^\prime}^{N_1,N_2} v_{12}(r_{ii^\prime}) \;.
	\label{hamiltonian}
\end{eqnarray}
Here, $a_z=\sqrt{\hbar/m\omega_z}$ is the oscillator length of the confining harmonic potential in the $z$ direction and $r_{ij}=|{\bf r}_i-{\bf r}_j|$ indicate interparticle distances with ${\bf r}_i$ and ${\bf r}_{i^\prime}$ referring to the first and second component of the mixture having number of particles, respectively, $N_1$ and $N_2$. Interaction potentials are modeled by repulsive hard spheres of diameter $a$ for $v(r)$ and by a pseudopotential with a negative scattering length $a_{12}$ for $v_{12}(r)$. Furthermore, we consider equal mass $m$ for the two components and equal particle numbers $N_1=N_2=N/2$. The simulations are performed using the PIMC algorithm described in Refs.~\cite{PhysRevLett.96.070601,condmat7020030}.
Both periodic boundary conditions (PBC) in the $x$-$y$ plane and a planar hard-wall circular confinement are implemented. Quasi-2D kinematic conditions are ensured by the temperature being a small fraction of the transverse confinement $k_BT=0.2\hbar\omega_z$ and by the small ratio of scattering to oscillator length $a/a_z=0.03162$ which is easily achievable in experiments with ultracold gases~\cite{PhysRevLett.99.040402,PhysRevA.95.013632,PhysRevLett.121.145301,PhysRevLett.127.023603}.

\begin{figure}[tb]
	\centering
	\includegraphics[width=0.94\columnwidth]{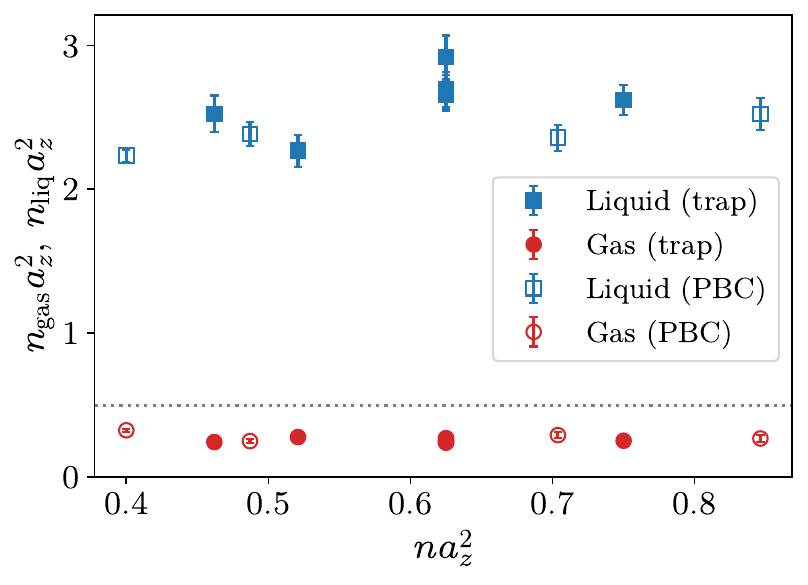}%
	\caption{
		Average densities within the liquid droplet $n_{\mathrm{liq}}a_z^2$ (blue squares) and in the gas region $n_{\mathrm{gas}}a_z^2$ (red circles), as a function of the global average density $na_z^2$. Open symbols correspond to the box with PBC, while full symbols correspond to the cylindrical hard-wall confinement. The other parameters are as in Fig.~\ref{fig1}. The dashed horizontal line indicates the critical density of the BKT transition for a homogeneous mixture without interspecies interactions.
	}
	\label{fig2}
\end{figure}

Results with PBC are shown in Fig.~\ref{fig1} [panels (a) and (b)]. The figure corresponds to a snapshot of particle positions projected onto the $x$-$y$ plane. The large system size allows us to clearly identify a region of high density (droplet) surrounded by a region of low density. By reducing the volume of the simulation cell the droplet occupies a larger portion of it while the densities of the inner and outer regions remain constant (see Fig.~\ref{fig2}). This behavior is characteristic of a liquid droplet in equilibrium with the surrounding gas. A similar behavior is observed for the mixture subject to a cylindrical hard-wall confinement [panel (c) in Fig.~\ref{fig1}]. An algorithm is devised to divide the $x$-$y$ plane of the simulation box into tiles and determine from snapshots of the projected particle positions the local 2D density $n$. By comparing the density in neighboring tiles we identify which tiles belong to the droplet, gas, or interface region. An example of this classification is shown in panels (a) and (b) of Fig.~\ref{fig1}. 
For the parameters of panels (a) and (b), the liquid (interface) region comprises approximately 10\% (15\%) and 53\% (27\%) of the total particles, respectively.
The obtained average densities of the liquid and gas tiles are reported in Fig.~\ref{fig2} as a function of the global average density both for PBC and circular wall configurations. We observe a distinctive jump from the gas density, $na_z^2\simeq 0.3$, to the liquid density, $na_z^2\simeq 2.5$, which is independent of the average global density in the simulation cell and of the geometry (PBC vs~circular). 
While the core liquid density does not depend on the droplet size, we find it varies with the scattering length ratio $a_{12}/a$ (see, e.g., Ref.~\cite{PhysRevA.109.013313}).
The almost $10\times$ increase of the density is associated with the first order gas to liquid transition and the discontinuity $\Delta n$ corresponds to the Isinglike scalar order parameter. We emphasize that in 2D the complex order parameter emerging from the breaking of the $U(1)$ symmetry is destroyed by thermal fluctuations according to the Mermin-Wagner theorem. Nonetheless, the scalar order parameter shown in Fig.~\ref{fig2} is well defined and is characteristic of attractive Bose mixtures close to the region of mean-field collapse. The situation is different in 3D droplets, where the gas-liquid transition is accompanied by BEC and the condensate density, being zero in the gas phase and finite in the liquid, accounts for the density discontinuity~\cite{PhysRevLett.131.173404}. We also notice that $n_{\mathrm{liq}}$ in Fig.~\ref{fig2} is significantly larger than the predicted critical density for the BKT transition corresponding to a homogeneous fluid in the absence of interspecies interactions, pointing toward superfluidity in the liquid state.

\begin{figure}[tb]
	\centering
	\includegraphics[width=0.98\columnwidth]{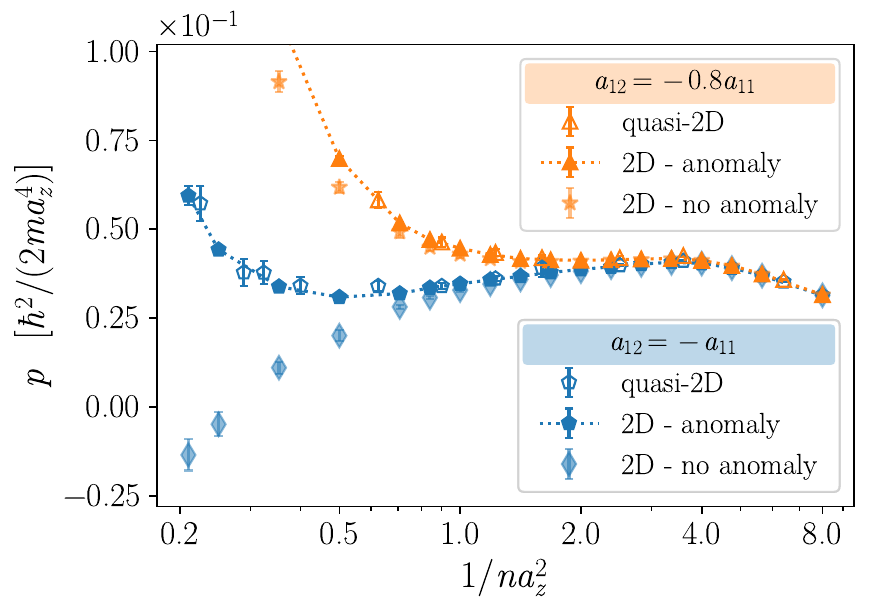}
	\caption{
		Pressure as a function of the inverse density for the quasi-2D mixture with scattering lengths $a_{12}=-0.8 a$ (orange open symbols) and $a_{12}=-a$ (blue open symbols). Temperature and intraspecies coupling constants are such that: $\lambda_T/a_z=5.605$ and $a=0.01995a_z$. The system size is $N_1+N_2=256$. Solid symbols refer to pure 2D simulations with matching parameters for temperature and couplings. In particular pale symbols correspond to simulations where 2D coupling strengths are constant and do not contain the logarithmic dependence on density.
	}
	\label{fig:quasi2d_comparison}
	\label{fig3}
\end{figure}

The gas-liquid coexistence is also investigated analyzing the behavior of the pressure vs.~inverse density. In Fig.~\ref{fig3} we report the results for two choices of the interspecies scattering length: in one case ($a_{12}=-0.8a$) the pressure is a monotonically increasing function with decreasing inverse density indicating a continuous crossover from a low-density to a high-density fluid without transition. On the contrary, in the other case, $a_{12}=-a$, the pressure exhibits an $S$-shape curve, typical of finite-size systems with a liquid-gas coexistence. Only in the thermodynamic limit one expects a horizontal pressure line. At higher density, the pressure increases again once the mixture is in the homogeneous liquid phase.

To gain understanding on the universality of the droplet formation and the observed behavior of the pressure, we perform simulations using a pure 2D model. The dimensionless 2D coupling constant for the intraspecies repulsion is defined as~\cite{RevModPhys.80.885, PhysRevA.64.012706,Pricoupenko_2007}
\begin{equation}
	g=\frac{g_0}{1+\frac{g_0}{4\pi}\log(A/na_z^2)} \;,
	\label{coupling}
\end{equation}
with logarithmic accuracy. Here $g_0=\sqrt{8\pi}a/a_z=0.1$, $A=0.2285$ is a numerical coefficient and $n$ is the total density of the mixture.
The $\log$ term in Eq.~\eqref{coupling} gives rise to the quantum anomaly breaking 2D scaling invariance.
An expression analogous to Eq.~\eqref{coupling} holds for the dimensionless interspecies coupling constant $g_{12}$ upon the substitution of $g_0$ with $(g_{12})_0 = \sqrt{8\pi}a_{12}/a_z$. A couple of remarks on Eq.~\eqref{coupling} are in order. First, the coupling dependence on the density exhibits different behaviors for repulsive and attractive interactions, specifically the former becomes stronger with increasing density, whereas the latter becomes weaker. Second, the 2D gas is assumed to be homogeneous. While this assumption is violated within the liquid-gas coexistence region, the mild logarithmic dependence allows one to employ Eq.\eqref{coupling} with the average density $n$ even in separated systems, provided the densities of the liquid and the gas are not too different.

In a PIMC simulation based on the pair-product ansatz~\cite{RevModPhys.67.279} interaction effects are included in the two-body density matrix
\begin{eqnarray}
	\rho_{\mathrm{rel}}({\bf r},{\bf r}^\prime,\tau)&=&\rho_{\mathrm{rel}}^0({\bf r},{\bf r}^\prime,\tau)
	+\frac{g}{8\pi} \int_0^\infty dk k e^{-\tau\hbar^2k^2/m}
	\nonumber\\
	&\times& \left[ J_0(kr)Y_0(kr^\prime)
		+Y_0(kr)J_0(kr^\prime)\right] \;,
	\label{densitymatrix}
\end{eqnarray}
where $\rho_{\mathrm{rel}}^0({\bf r},{\bf r}^\prime,\tau)=\frac{m}{4\pi\tau\hbar^2}e^{-({\bf r}-{\bf r}^\prime)^2m/(4\tau\hbar^2)}$ is the noninteracting pair density matrix in 2D as a function of the particle positions ${\bf r}$ and ${\bf r}^\prime$ in the $x$-$y$ plane and of the inverse temperature step $\tau$. Furthermore, $g$ is the dimensionless 2D coupling strength given in Eq.~(\ref{coupling}). The functions $J_0(x)$ and $Y_0(x)$ in Eq.~(\ref{densitymatrix}) are Bessel functions of the first and second kind depending on the products $kr$ and $kr^\prime$ of the relative wave vector $k$ and the modulus of position vectors. Equation (\ref{densitymatrix}) refers to intraspecies interactions between pairs of particles in the same component. For interspecies interactions we use a similar expression for the corresponding two-body density matrix where the coordinates ${\bf r}$ and ${\bf r}^\prime$ indicate pairs of particles in different components and $g$ is replaced by $g_{12}$. The above expression for the pair density matrix has been already used in PIMC studies of 2D Bose gases~\cite{PhysRevLett.111.050406}. It accounts only for the spherical ($\ell=0$) contribution to scattering and holds to lowest order in the coupling constant~\cite{PhysRevLett.126.110401}.

The comparison between quasi-2D and pure 2D models is shown in Fig.~\ref{fig3}. The coupling strengths of the 2D simulations are matched to the 3D values using Eq.~(\ref{coupling}) and the 2D density $n\lambda_T^2$ matches the value $na_z^2$ with $\lambda_T=5.605 a_z$ where $\lambda_T=\sqrt{2\pi\hbar^2/mk_BT}$. The nice agreement of the two results for the pressure, both at $a_{12}=-0.8a$ and $a_{12}=-a$, indicates the reliability of the pure 2D model.
Crucially, we find that the quantum anomaly term is essential for reproducing the quasi-2D results of the pressure in the regime of high density and, in particular, for the gas-liquid coexistence region in the balanced case $a_{12}=-a$. In Fig.~\ref{fig3}, we demonstrate this by neglecting the $\log$ term of the density in Eq.~\eqref{coupling}, i.e., imposing constant values for the coupling strengths $g=g_0$ and $g_{12}=(g_{12})_0$.

\begin{figure}
	\centering
	\includegraphics[width=0.98\columnwidth]{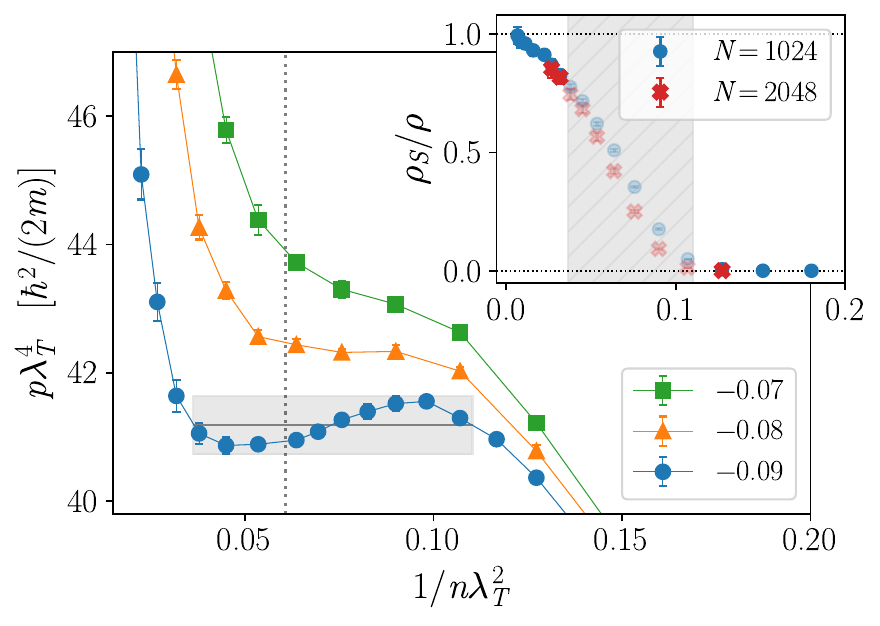}
	\caption{Pressure as a function of the inverse density for the pure 2D model for three values of the interspecies coupling strength $(g_{12})_0$ reported in the legend. The intraspecies coupling is $g_0=0.1$ and $N_1 + N_2=1024$. The solid lines are a guide to the eye and the horizontal gray band gives the coexistence region as obtained from the Maxwell construction~\cite{huang2008statistical}. The vertical dotted line represents the expected location of the BKT transition for the noninteracting mixture. \emph{Inset}: the superfluid fraction as computed from eq.~\eqref{eq:superfluid_fraction}, vs the inverse density $1/n\lambda_T^2$, for the $(g_{12})_0=-0.09$ case. Red crosses represent the data for $N_1 + N_2=2048$. The shaded area is the coexistence region, where the winding formula is not applicable.}
	\label{fig:press2d_couplings}
	\label{fig4}
\end{figure}

Having established the applicability of the pure 2D model, we use it to explore different values of the interspecies coupling strength as well as the superfluid response provided by the winding number estimator~\cite{PhysRevB.36.8343,PhysRevB.39.2084}
\begin{equation}
	\frac{\rho_S}{\rho}=\frac{mk_BT}{2N\hbar^2}\langle {\bf W}^2\rangle \;,
	\label{eq:superfluid_fraction}
\end{equation}
where ${\bf W}=\sum_{i=1}^N[{\bf r}_i(\beta)-{\bf r}_i(0)]$ in terms of the initial ($\tau=0$) and final ($\tau=\beta$) time-step position of all the particles in the mixture. The results are shown in Fig.~\ref{fig4}, where inverse density and pressure are plotted in natural units of the pure 2D system. In these units, pressure lines exhibit only a very weak, logarithmic, temperature dependence due to the quantum anomaly. In fact, they are mostly determined by the values of the coupling strengths $g_0$ and $(g_{12})_0$. The critical point of the gas-liquid transition occurs for a value of $(g_{12})_0$ slightly lower than $(g_{12})_0=-0.08$. For $(g_{12})_0>-0.08$ the pressure increases monotonically without featuring a density jump, while for stronger attraction a gas-liquid coexistence line can be drawn using the Maxwell construction.
Neglecting interspecies interactions, the mixture is expected to cross the BKT transition at $n/2=n_{\mathrm{BKT}}$, with $n_{\mathrm{BKT}}$ the critical density of a single-component gas $n_{\mathrm{BKT}}\lambda_T^2=\log(\xi/g_0)$ where $\xi\simeq380$~\cite{PhysRevLett.87.270402}.
For systems featuring the gas-liquid transition, the corresponding inverse density lies within the coexistence region, as shown in Fig.~\ref{fig4} for the $(g_{12})_0=-0.09$ case, indicating that the transition is first order and takes place from a normal gas to a superfluid liquid.
This result is confirmed by the calculation of the superfluid component defined in Eq.~\eqref{eq:superfluid_fraction} and reported in the inset of Fig.~\ref{fig4}. This calculation is well defined in the homogeneous gas and liquid phases, while it has no meaning in the coexistence region where the mixture consists of liquid droplets separated by gas. As a consequence of this granularity, the results of $\rho_S/\rho$ in the coexistence region exhibit a significant size dependence. Outside this region we find that $\rho_S/\rho$ is compatible with zero in the gas phase and jumps to $\rho_S/\rho\simeq 0.8$ in the liquid phase.

In conclusion, we have carried out a thorough analysis of attractive Bose mixtures in planar geometries with strong transverse confinement using exact PIMC methods. We observe the formation of liquid droplets in equilibrium with the gas in regimes of 2D densities easily achievable in current experiments. The first-order gas to liquid transition is accompanied by a tenfold increase of the density. The use of a pure 2D model allows us to establish that the results are universal in terms of the 2D coupling strengths and, in particular, depend crucially on the quantum scale anomaly. Furthermore, within the pure 2D model, we analyze  how the first-order transition terminates for increasing interspecies coupling strengths, as well as how the superfluid fraction jumps at the first-order gas to liquid transition.\\

\begin{acknowledgments}
    \emph{Acknowledgments}---S.P. and G.S. acknowledge support from the Italian Ministry of University and Research under the PRIN2022 project ``Hybrid algorithms for quantum simulators'' No. 2022H77XB7, and from CINECA for the availability of high performance computing resources and support under the ISCRA initiative.
	S.P. also acknowledges support from the PNRR MUR Project No. PE0000023-NQSTI and from the EuroHPC Joint Undertaking, for awarding access to the EuroHPC supercomputer LUMI, hosted by CSC (Finland) and the LUMI consortium through a EuroHPC Benchmark Access call.
    G.S. and S.G. acknowledge funding from the Provincia Autonoma di Trento.
	S.G. also acknowledges support from ICSC – Centro Nazionale di Ricerca in HPC, Big Data and Quantum Computing, funded by the European Union under NextGenerationEU. Views and opinions expressed are, however, those of the author(s) only and do not necessarily reflect those of the European Union or The European Research Executive Agency. Neither the European Union nor the granting authority can be held responsible for them.
\end{acknowledgments}

\bibliography{bibliography}

\end{document}